\documentclass[useAMS, usenatbib]{mn2e}
\usepackage{times}
\usepackage{amsmath}
\usepackage{xspace}
\usepackage{psfig}
\usepackage{defs}

\def\eq{equation}
\def\fig{Fig.}
\def\figs{Figs.}
\def\tab{table}

\def\cf{{\it cf.}}
\def\ie{{\it i.e.}}
\def\eg{{\it e.g.}}
\def\ltsima{$\; \buildrel < \over \sim \;$}
\def\simlt{\lower.5ex\hbox{\ltsima}}
\def\gtsima{$\; \buildrel > \over \sim \;$}
\def\simgt{\lower.5ex\hbox{\gtsima}}
\def\fesc{{$\langle f_{\rm esc}\rangle$}\xspace}
\def\h2{H$_2$\xspace}
\def\H2{H$_2$\xspace}
\def\m{$^{-1}$\xspace}
\def\mm{$^{-2}$\xspace}
\def\mmm{$^{-3}$\xspace}
\def\pp{$^2$\xspace}

\def\ion#1#2{\text{#1\,\sc #2}}
\def\HI{{\ion{H}{i} }}
\def\HII{{\ion{H}{ii} }}
\def\GI{{\ion{He}{i} }}
\def\GII{{\ion{He}{ii} }}
\def\GIII{{\ion{He}{iii} }}

\def\popII{Population~II\xspace}
\def\pop3{Population~III\xspace}

\def\p3{``small-halo''\xspace}
\def\pp3{``Small-halo''\xspace}

\def\Mpc{h$^{-1}$ Mpc\xspace}

\def\pc{$h^{-1}$~pc\xspace}
\def\Ms{$h^{-1}$~M$_\odot$\xspace}

\def\lya{Lyman-$\alpha$\xspace}
\def\lyb{Lyman-$\beta$\xspace}

\def\taue{$\tau_{\rm e}$\xspace}
\def\sig8{$\sigma_8$\xspace}

\def\xe{$\langle x_{\rm e} \rangle$\xspace}
\def\bi{\begin{itemize}}
\def\ei{\end{itemize}}

\title[X-ray Preionisation. II]{X-ray Preionisation Powered by Accretion
  on the First Black Holes. II: Cosmological Simulations and Observational Signatures}

\author[M. Ricotti, J.P. Ostriker and N.Y. Gnedin]
{Massimo Ricotti$^1$, Jeremiah P. Ostriker$^1$ and Nickolay Y. Gnedin$^2$\\ 
$^1$ Institute of Astronomy, Madingley Road, Cambridge CB3 0HA, UK\\
$^2$ Center for Astrophysics and Space Astronomy, University of Colorado, Campus Box 389, Boulder, CO 80309, USA\\
ricotti@ast.cam.ac.uk, jpo@ast.cam.ac.uk, gnedin@casa.colorado.edu}
\date{Accepted ---. Received ---; in original form 10 December 2002}
\pagerange{\pageref{firstpage}--\pageref{lastpage}}
\pubyear{2002}
\begin{document}
\maketitle
\label{firstpage}

\begin{abstract}
  We use numerical simulations of a cosmological volume to study the
  X-ray ionisation and heating of the intergalactic medium by an early
  population of accreting black holes. By considering theoretical
  limits on the accretion rate and observational constraints from the
  X-ray background and faint X-ray source counts, we find that the
  maximum value of the optical depth to Thompson scattering that can
  be produced using these models is \taue$\simeq 0.17$, in agreement
  with previous semianalytic results.  The redshifted soft X-ray
  background produced by these early sources is important in producing
  a fully ionised atomic hydrogen in the low density intergalactic
  medium before stellar reionisation at redshift $z \sim 6-7$. As a
  result stellar reionisation is characterised by an almost
  instantaneous ``overlap phase'' of \HII regions. The background also
  produces a second \GII reionisation at about redshift three and
  maintains the temperature of the intergalactic medium at about
  10,000 K even at low redshifts.
 
  If the spectral energy distribution of these sources has a
  non-negligible high energy power-law component, the luminosity in
  the soft X-ray band of the ``typical'' galaxies hosting
  intermediate-mass accreting black holes is maximum at $z \sim 15$
  and is about one or two orders of magnitude below the sensitivity
  limit of the Chandra deep field. We find that about a thousand of
  these sources may be present per square arcmin of the sky, producing
  potentially detectable fluctuations. We also estimate that a few
  rare objects, not present in our small simulated volume, could be
  luminous enough to be visible in the Chandra deep field.  XEUS and
  Constellation-X satellites will be able to detect more of these
  sources that, if radio loud, could be used to study the 21 cm forest
  in absorption.
  
  A signature of an early X-ray preionisation is the production of
  secondary CMB anisotropies on small angular scales ($<1$ arcmin).
  We find that in these models the power spectrum of temperature
  fluctuations increases with decreasing angular scale ($\Delta T \sim
  16 \mu$K at $\sim 1$ arcsec scales), while for stellar reionisation
  scenarios the power decreases on smaller scales.  We also show that
  the redshifted 21 cm radiation from neutral hydrogen can be
  marginally detected in emission at redshifts $7<z<12$. At a redshift
  of about $z \sim 30$ a stronger and narrower (in redshift space)
  signal in absorption against the CMB, that is peculiar to these
  models, could be detectable.
\end{abstract}

\section{Introduction}

The WMAP satellite measured an optical depth to Thompson scattering of
the IGM \taue$\simeq 0.17 \pm 0.04$ \citep{Kogut:03}. The result,
implies a much earlier start of reionisation with respect to the
redshift when reionisation was completed (at $z_{\rm rei} \sim 6$, as
estimated from the absorption spectra of high redshift quasars). This
has also been interpreted as one of the stronger pieces of evidence
for the importance of zero-metallicity (\pop3) stars in the early
universe \citep[\eg,][]{Cen:03b, WyitheL:03, Somerville:03,
  CiardiFW:03, Sokasian:03, ChiuFO:03} and of the formation of the
first small mass galaxies (with $M_{dm} \sim 10^6-10^8$ M$_\odot$), a
topic that is still under debate \citep[\eg,][]{HaimanRL:97,
  RicottiGS:01, RicottiGSb:02, Machacek:03}.  In a previous paper
(paper~I), \cite{RicottiOI:03} have argued that, even assuming the
most favorable properties for \pop3, it is difficult or impossible to
produce the measured value of \taue using UV radiation from \pop3
stars. The main problems with this scenario are the negative feedback
from ionising radiation, mechanical feedback from SN explosions and
the contamination of high density regions by detritus from the same
stars which produce the ionising radiation.  Taking into account the
metal enrichment from SNe and pair instability SNe the \pop3 epoch
turns out to be so short lived that \pop3 stars will never be able to
complete the reionisation of the IGM.  But if a significant fraction
of \pop3 stars, instead of exploding as SNe, implode into black holes
(BHs), then the negative feedback on star formation is much reduced
and the UV radiation produced by thermonuclear reactions in the first
stars might be important for reionisation. In \cite{RicottiO:03}
(paper~IIa) we have shown that in such a scenario, the X-ray
background produced by accretion onto stellar mass seed black holes
would be more effective in producing the large \taue measured by WMAP
than the UV radiation from the parent stars. \cite{MadauR:03} have
also investigated a scenario in which accretion on intermediate-mass
black holes produces the large optical depth to Thompson scattering
measured by WMAP. In their paper they focus is on the physics of black
hole mergers and accretion onto seed black holes. Their paper is
complementary to our studies that, instead, focus on radiative transfer
processes.  Before the WMAP measurement of a large \taue, the effect
of X-ray preionisation has been investigated by \cite{Oh:00} and
\cite{Venkatesan:01}. The values of \taue that they found were smaller
since the \HI ionisation was produced by secondary photoelectrons that
can ionise the gas to a maximum of 10\% ionisation fraction. We find
larger values of \taue, consistent with WMAP, because of the
additional ionisations produced by the redshifted soft X-ray
background.
 
\begin{table*}
\centering
\caption{Input parameters for hydrodynamic simulations with radiative transfer.\label{tab:1}}
\begin{tabular*}{15. cm}[]{ll|cccccccccc}
\# & {RUN} & {$N_{\rm box}$} & {$L_{\rm box}$} & {Mass Res.} & {Res.} & {$g_\nu$} & 
{$\epsilon_*$}&{$\epsilon_{\rm UV}$\fesc} &{$\epsilon_{\rm qso}$} &
$z_{\rm off}$ & $F_{\rm IMF}$ \\
{} & {} & {} & {$h^{-1}$ Mpc} & {$h^{-1}$ M$_\odot$} & {$h^{-1}$ pc} & {} &
{} & {} & {} & {} & {}\\
\hline
1 & M-PIS & 128 & 1.0  & $3.94\times 10^4$ & 488 & Pop~III & 0.1 & 
$2\times 10^{-3}$ & $2\times 10^{-3}$ & 8 & 7\\
2 & M-SN1 & 128 & 1.0  & $3.94\times 10^4$ & 488 & Pop~III & 0.1 & 
$3\times 10^{-4}$ & $2\times 10^{-3}$ & 10 & 1\\
3 & M-SN2 & 128 & 1.0  & $3.94\times 10^4$ & 488 & Pop~III & 0.1 &
$3\times 10^{-5}$ & $2\times 10^{-3}$ & 11 & 0.1\\
4 & M-BH & 128 & 1.0  & $3.94\times 10^4$ & 488 & Pop~III & 0.1 & 
$3\times 10^{-5}$  & see Eq.~\ref{eq:X3} & 13 & 0.1\\
\end{tabular*}
\flushleft
{Parameter description. {\em Numerical parameters:}
   $N_{\rm box}^3$ is the number of grid cells, $L_{\rm box}$ is the box size
   in comoving \Mpc. {\em Physical parameters:} $g_\nu$ is the
   normalised SED (\pop3). The meaning of the other parameters is
   explained in the text (\S~\ref{ssec:code}).}
\end{table*}

In this paper we complement the semianalytic results presented in
paper~IIa by using hydrodynamic cosmological simulations that include a
recipe for star formation and 3D radiative transfer for hydrogen and
helium ionising radiation. An approximate solution of the radiative
transfer equations \citep{GnedinA:01} is used to speed up the
calculations and make the coupling of the radiative transfer and
hydrodynamic equations computationally feasible. In addition,
radiative transfer for the optically thin X-ray radiation and \H2
dissociating radiation is solved exactly. Line radiative transfer in
the \H2 Lyman-Werner bands ($11.3 < h_P\nu <13.6$ eV) is also solved
exactly for the volume averaged component of the radiation field.
Using this method we are able to simulate local and global radiative
feedback effects of galaxy formation on cosmological scales.

The results of the cosmological simulations are used to predict the
distinctive observational signatures of the X-ray preionisation
scenario compared to stellar reionisation models.  We find that, if
this scenario is correct, a new population of X-ray sources that do
not have optical counterparts may be detected in the Chandra deep
field. Interestingly, these sources, that might be high redshift AGNs
but also galaxies hosting a large number of bright Ultra-luminous
X-ray sources (ULXs), might have been already observed
\citep{Koekemoer:03}.

Simulations of a cosmological volume allow us to construct maps for
the secondary CMB anisotropies produced after recombination by the
first sources of ionising radiation. We also use the simulation
outputs and the semianalytic models presented in paper~IIa to
calculate the redshifted 21 cm radiation from the IGM prior to
reionisation in emission or absorption against the CMB. In the X-ray
preionisation scenario the IGM is only partially ionised at redshift
$z>z_{\rm rei} \simeq 6-7$.  Therefore, the 21cm signal from the IGM
at $z<12$ can be detected even though it is rather weak. Note that, in
stellar reionisation scenarios that can produce the optical depth to
Thompson scattering observed by WMAP, the expected 21cm signal is
instead undetectable.  Finally we estimate the additional high energy
background due to the postulated population of high redshift X-ray
sources and its subsequent signatures [$(10 \pm 5)$\% in the 2-50 keV
bands].

This paper is organised as follows: in \S~\ref{sec:res_cs} we show the
results of cosmological simulations of X-ray preionisation by mass
accretion on seed BHs. In \S~\ref{sec:lum} we estimate the number of
detectable X-ray point sources at high redshift. In \S~\ref{sec:21cm}
we calculate the redshifted 21cm signal for one of our simulations in
absorption/emission against the CMB. In \S~\ref{sec:cmb} we compute
the amplitude of the power spectrum of CMB secondary anisotropies on
scales of a few arcmin produced by X-ray preionisation and stellar
reionisation scenarios. We summarise the results in \S~\ref{sec:conc}
and we discuss the observational signatures of the X-ray preionisation
scenario when compared to stellar reionisation scenarios.

\section{Cosmological Simulations}\label{sec:res_cs}

In this section we show the results of cosmological simulations
including radiative feedback effects and feedback from SN explosions.
The code has been implemented and used extensively to study the
formation of the first galaxies.  In this study we have run four new
simulations to study the effects of SN explosions and an early X-ray
background.  The simulations were run on COSMOS, an SGI Origin 38000
in DAMPT at Cambridge University.

We adopt a concordance $\Lambda$CDM cosmological model with parameters
consistent with the analyses of \cite{Spergel:03} and
\cite{Tegmark:03}: $\Omega_{\rm m} = 0.3$, $\Omega_\Lambda= 0.7$,
$h=0.7$ and $\Omega_{\rm b} = 0.04$.  The initial spectrum of
perturbations has $\sigma_8=0.91$ and $n_{\rm s}=1$.  The box size is
$L_{box}=1$ \Mpc (comoving) and the grid has $N_{box}^3=128^3$ cells.
We achieve a maximum mass resolution of $M_{DM}=3.94\times 10^4$ \Ms
and spatial resolution of $488$ \pc (comoving).  We fully resolve the
star formation in objects within the mass range $5 \times 10^5~{\rm
  M}_\odot \simlt M_{DM} \simlt 10^9~{\rm M}_\odot$. All simulations
start at $z = 100$ and end at $z \approx 8$. After this redshift we
stop the simulations because the simulated volume ceases to be a
statistically representative fraction of the universe.

\subsection{The Code}\label{ssec:code}

The simulations were performed with the ``Softened Lagrangian
Hydrodynamics'' (SLH-P$^3$M) code described in detail in
\cite*{Gnedin:95}. The cosmological simulation evolves collisionless
DM particles, gas, ``star-particles'' and the radiation field in four
frequency bands: optically thin radiation, \HI, \GI and \GII ionising
radiation. The radiative transfer is treated self-consistently (\ie,
coupled with the gas dynamics and star formation) using the OTVET
approximation \citep{GnedinA:01}. Star particles and BHs are
formed as per \eq~(\ref{eq:sfr}) in each resolution element that sinks
below the spatial resolution of the code. The code adopts a deformable
mesh to achieve higher resolution in the dense filaments of the large
scale structure. We solve the line radiative transfer in the \H2
Lyman-Werner bands for the background radiation, we include the effect
of secondary ionisation of H and He by X-rays, heating by Ly$\alpha$
scattering, detailed \H2 chemistry and cooling, and the absorbed
stellar energy distribution (SED) of the sources that depends on the
mean UV escape fraction \citep{RicottiGSa:02}.  Here we also include
the effect of SN explosions using the method in \cite{Gnedin:98a}.  In
\cite{RicottiGSa:02} we discussed extensively the details of the code
and the physics included in the simulation, focusing on simulations of
the first galaxies. We also performed numerical convergence studies
that are especially crucial in this case. High mass resolution is
needed because the objects that we want to resolve have masses
$10^5~{\rm M}_\odot \simlt M_{\rm dm} \simlt 10^8~{\rm M}_\odot$.
Moreover, the box size has to be large enough in order to include at
least a few of the rare first objects. The first small-mass galaxies
will form at $z~\sim 30-40$ from $3\sigma$ or more rare density
perturbations; the first normal galaxies with $M_{\rm dm} \simgt 5
\times 10^8$ M$_\odot$ form at $z \simgt 15$, also from $3\sigma$
perturbations.  Below, we summarise the meaning of the free parameters
in the simulations.
\begin{figure*}
\centerline{\psfig{figure=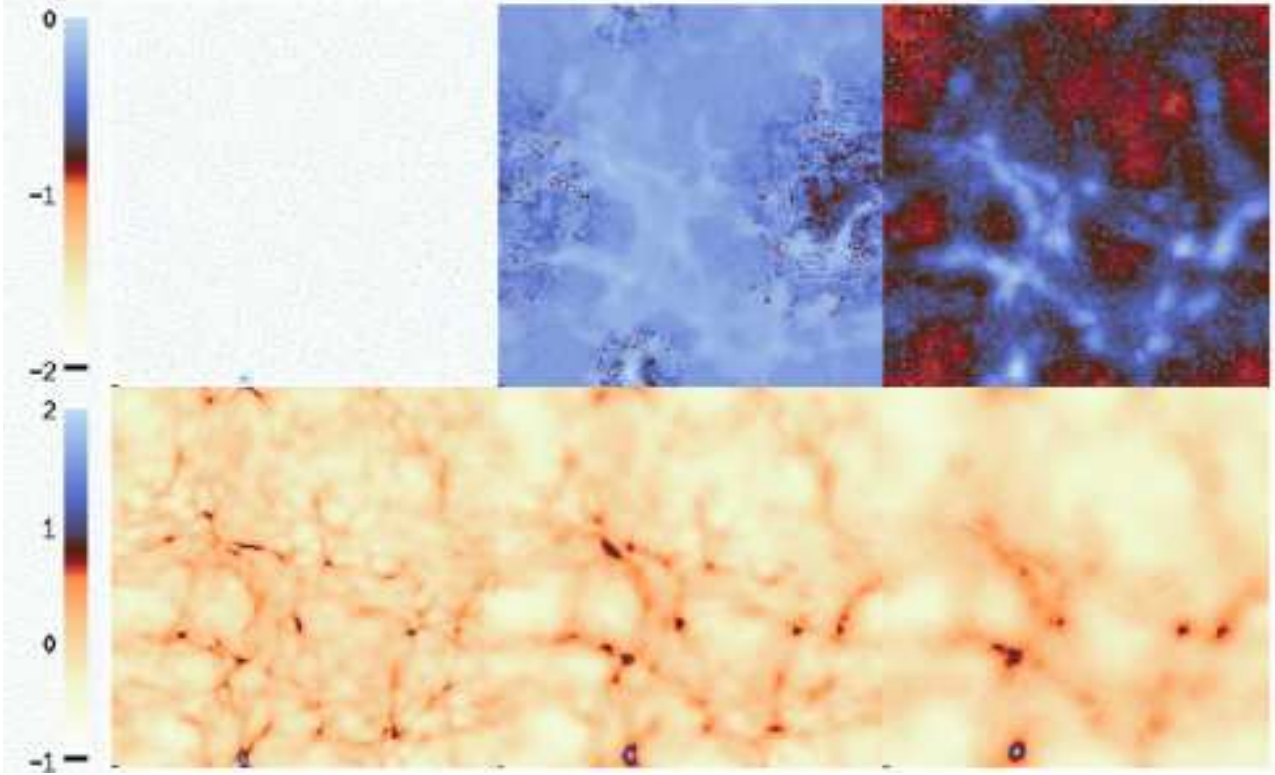,width=17cm}}
\caption{\label{fig:slice} Slices through the volume of the simulation 
  M-BH (\cf, \tab~\ref{tab:1}) at redshift $z =19, 12.5$ and $9$
  from left to right. The top panels show the hydrogen ionisation
  fraction (in logarithmic scale) and the bottom panels the gas
  overdensity (in logarithmic scale). Note the ionisation fraction in
  the voids is larger than in the denser filaments. The overdense gas
  is smoothed on the filtering scale (\ie, time averaged Jeans length)
  by the X-ray reheating that precedes the IGM partial preionisation.}
\end{figure*}

\begin{itemize}
\item $\epsilon_*$: Efficiency of star formation for the adopted star
  formation law,
\begin{equation}
{d\rho_* \over dt} = \epsilon_* {\rho_{\rm gas} \over t_*},
\label{eq:sfr}
\end{equation}
where $\rho_*$ and $\rho_{\rm gas}$ are the stellar and gas density,
respectively.  $t_*$ is the maximum between the dynamical and cooling
time.
\item $\epsilon_{\rm UV}$: Energy in ionising photons emitted by stars
  per rest mass energy of H atoms ($m_{\rm H} c^2$) transformed into
  stars. This parameter depends on the IMF and stellar metallicity. We
  use \pop3 stars SED.
\item \fesc: escape fraction of ionising photons from a cell. It is
  resolution dependent.
\item $\epsilon_{\rm qso}$: Energy in ionising photons emitted by
  quasars per rest mass energy of H atoms ($m_{\rm H} c^2$)
  transformed into stars. This means that, if $\epsilon_{\rm qso}={\rm
    const}$, the total UV luminosity from quasars is proportional to
  that from massive stars.  We use the template spectrum given in
  \S~3.1 of paper~IIa. The spectrum is a double power law with a soft
  X-ray bump produced by absorption of UV photons by the obscuring
  torus and the ISM around the BH (in particular the absorption is
  produced by an \HI column density of $10^{19}$ cm\mm and He with
  ionisation fractions $x_\GI/x_\HI=0.1$, $x_\GII/x_\HI=0.1$).  In
  paper~I we mentioned that our spectrum is similar to the template
  spectrum derived by \cite{SazonovO:04}. This statement is actually
  misleading since our spectrum is softer, having a cutoff at about
  100 eV instead of 1~keV. We will discuss in \S~\ref{ssec:sed} the
  importance of soft X-rays and we will quantify how the cutoff energy
  in the quasar spectrum affects our results.  The energy density in
  the X-ray bands is about $\beta=20$\% of the total and 40 \% of the
  energy of H ionising radiation ( $\epsilon_{\rm X}=0.4 \epsilon_{\rm
    qso}$, because of the local absorption of UV photons).  The
  accretion rate at the Eddington limit with efficiency
  $\epsilon=0.2$, defined in \S~3 of paper~IIa, is given by $0.2 \beta
  \dot \rho_{\rm ac}= \epsilon_{\rm X} \dot \rho_{*}$, where $\dot
  \rho_*$ is the star formation rate and $\dot \rho _{\rm ac}$ is the
  accretion rate onto BHs.
\item $z_{\rm off}$: redshift at which the X-ray emission by quasars
  turns off. We assume that $\epsilon_{\rm qso}=0$ at $z<z_{\rm off}$.
\item $F_{\rm IMF}$: This parameter is proportional to the mean
  metallicity yield and energy input by SN explosions of the stellar
  population. We have $F_{\rm IMF}=1$ for \popII stars with a Salpeter IMF.
  But \pop3 stars could have $F_{\rm IMF}>1$ if the IMF is top-heavy
  and pair-instability SNe are dominant. If the IMF is dominated by
  subluminous BH forming SNe, or by stars with $M_* > 260$ M$_\odot$ that
  collapse directly into BHs without exploding as SNe (see
  \S~3 of paper~IIa), we have $F_{\rm IMF}<1$.
\end{itemize}
\begin{figure*}
\centerline{\psfig{figure=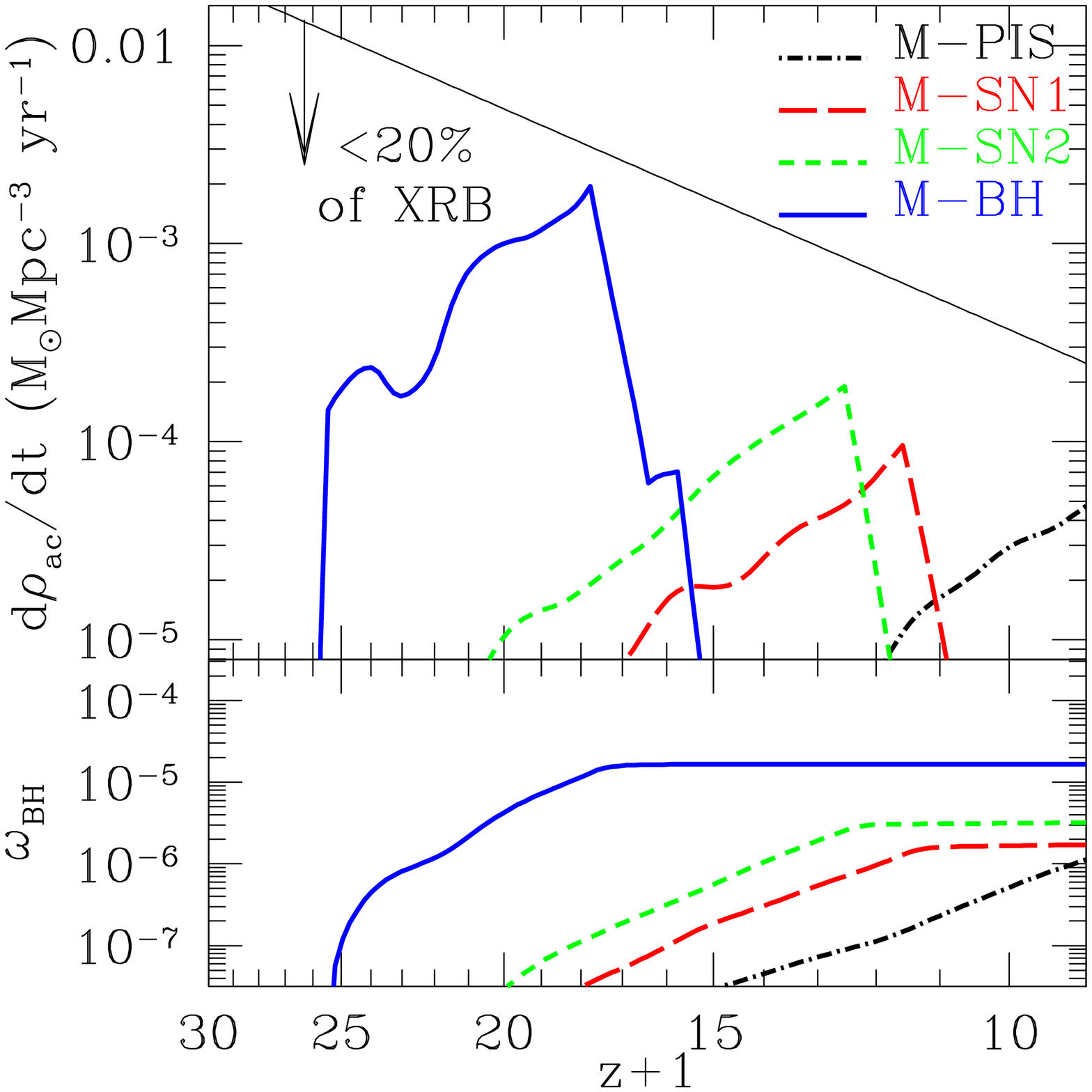,width=8.5cm}
\psfig{figure=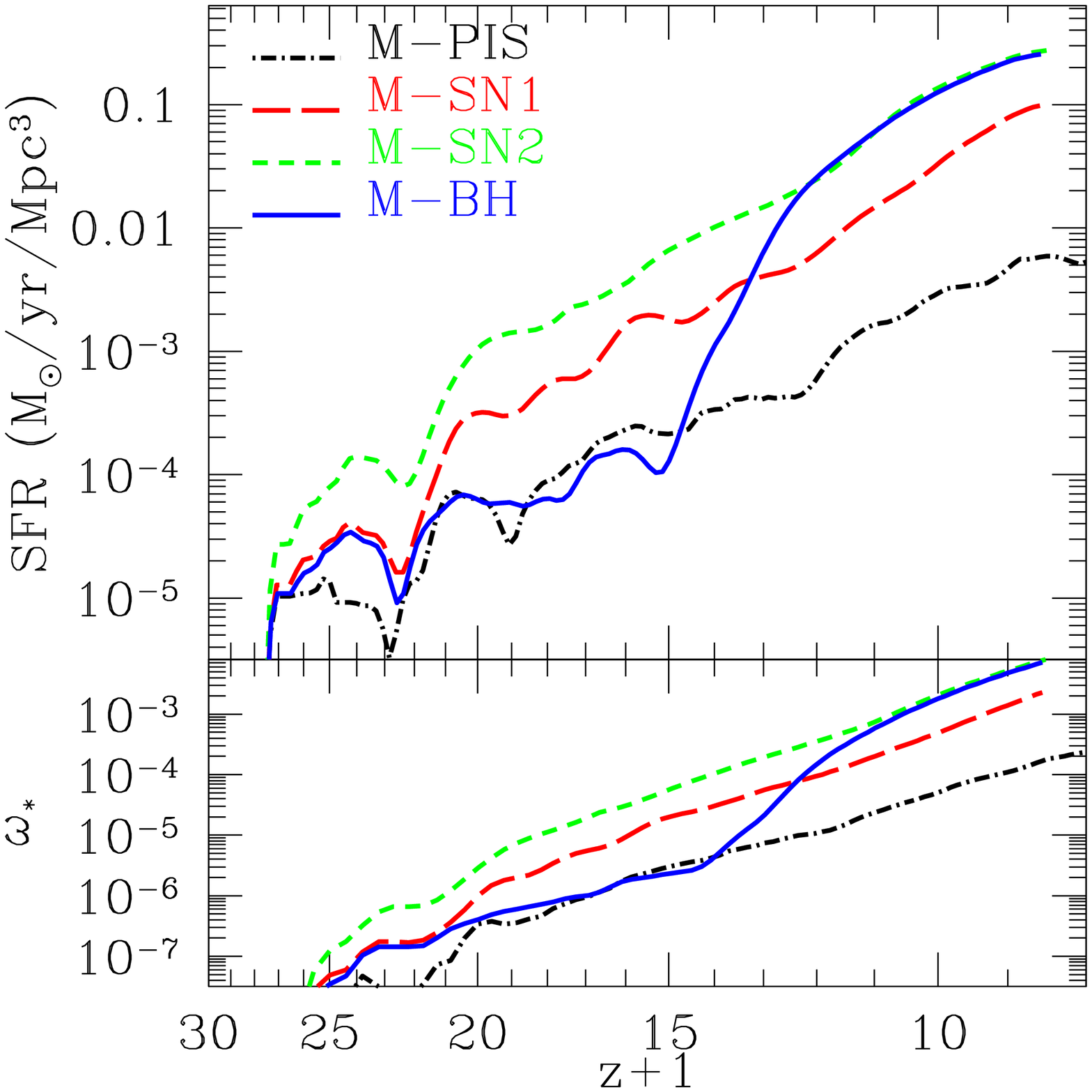,width=8.5cm}}
\caption{\label{fig:srei1} (left) Black hole accretion history for the
  four models in \tab~\ref{tab:1}. (right) Star formation history for
  the four models in \tab~\ref{tab:1}. The simulations differ in the
  efficiency of X-ray emission (duty cycle), IMF and \fesc. The solid
  line shows the upper limit of the accretion rate for the SED we have
  adopted. If the global accretion rate on seed BHs is below the solid
  line, their contribution to the observed X-ray background at $z=0$
  (at about 10 keV) is less than 20\%.}
\end{figure*}

We have run four simulations that differ mainly in the epoch at which
the early accretion onto seed BHs takes place (\cf, \tab~\ref{tab:1}).
In paper~IIa we have studied three models: an early preionisation
model (with preionisation starting at redshift $z_{\rm pre} \sim 25$),
an intermediate preionisation model ($z_{\rm pre} \sim 20$) and late
preionisation model ($z_{\rm pre} \sim 15$).  Due to the limited
volume and resolution of our simulations, the first sources form at
redshift $z \sim 27$, later than in the semianalytic models that do
not suffer of these limitations. The radiation background from the
first sources (\cf, \fig~\ref{fig:sback}) builds up to relevant values
only about one Hubble time after the formation of the first source,
corresponding to a redshift $z \sim 20$.  At this redshift the
ionisation fraction of the IGM will start increasing.  For this
reason, using numerical simulations, we cannot simulate the early
preionisation model presented in paper~IIa.  In order to simulate
earlier object formation, we would need simulations with comparable
mass resolution to the present but of a larger volume of the universe.
In such a simulation the first galaxies would have formed at redshift
$z \sim 40$ from rare (\ie, $5\sigma$) peaks of the Gaussian
perturbations in the density field, allowing the earlier production of
seed BHs and preionisation by the X-ray background.

In the first three simulations in \tab~\ref{tab:1} (M-PIS, M-SN1 and
M-SN2) we have assumed a step function for the efficiency of X-ray
emission $\epsilon_{\rm qso}$ (\ie, $\epsilon_{\rm qso}=0.002$ at
$z>z_{\rm off}$ and zero afterwards). This means that in the first
three simulations we assume that the local black hole accretion rate
is proportional to the local SFR and we explore the effects of
changing the stellar IMF and \fesc.  In the simulation M-PIS the IMF
is top-heavy, \fesc$=1$ and the energy input from SN explosions is 7
times larger than for a Salpeter IMF. The parameters of this first
simulation describe a scenario in which \pop3 stars are supermassive
and a substantial fraction of them explodes as pair-instability
supernovae. In the simulation M-SN1 we assume a Salpeter IMF and
\fesc$=1$.  The parameters of this simulation are also consistent with
a top-heavy IMF combined with \fesc$ \ll 1$. The simulation M-SN2 is
the same as M-SN1 but has \fesc$=10$\% and the effect of SN explosions
is $10$ times smaller. The parameters of this simulation are also
consistent with a mildly top-heavy IMF combined with \fesc$< 10$\% and
subluminous or negligible SN explosions. This assumption is justified
by the fact that zero-metallicity stars, if more massive than $\sim
260$ M$_\odot$, may collapse directly into BHs without exploding as
SN. Another possibility is that the energy of SN explosions in
zero-metallicity stars with masses $\simlt 100$ M$_\odot$, is smaller
than the canonical value $E=10^{51}$ ergs \citep{UmedaN:03}.

The fourth simulation (M-BH) is the same as M-SN2 but has larger and
time-dependent efficiency of X-ray emission, $\epsilon_{\rm qso}$.
The X-ray emissivity is similar to the one used in the semianalytic
model M3 (the intermediate preionisation model) of paper~IIa and
it is physically motivated in \S~3 of that paper.  The time dependent
function for $\epsilon_{\rm qso}$ is given by
\begin{equation}
\epsilon_{\rm qso}=0.2 \times
\begin{cases}
\exp[(39/(1+z))^{1.5}]~\text{if $z>17$}\\
0.0016((1+z)/15)^{20}~\text{if $z<17$}.
\end{cases}
\label{eq:X3}
\end{equation}
The efficiency $\epsilon_{\rm qso}$ has a maximum at $z=17$ and then
quickly decreases.  This is the most interesting simulation because
the function $\epsilon_{\rm qso}$ is a fit to a realistic model for
accretion and because, as shown in the next section, produces a value
of the optical depth in agreement with th WMAP measures (\cf,
\S~\ref{ssec:res}).

\subsection{Results}\label{ssec:res}

The soft X-ray and hard UV flux from the first accreting BHs are
initially the most efficient in ionising the intergalactic gas in the
immediate vicinity of each source. Thus the topology of IGM ionisation
is initially characterised by small Str\"omgren spheres around the
most luminous sources. But, due to the long mean free path of X-ray
photons, the background produced by distant sources quickly dominates
the ionisation rate and the IGM becomes partially ionised almost
uniformly. Due to the negligible recombination rate, the voids have a
larger fractional ionisation than the denser filaments. The evolution
of the topology of preionisation is shown in the top panels of
\fig~\ref{fig:slice} where we show the neutral fraction of the IGM in
a slice trough the simulation number four in \tab~\ref{tab:1} at
redshift $z=19, 12.5$ and $9$. The bottom panels show the gas
overdensity in the same simulation. It is evident from the smooth
appearance of the filamentary structure that the X-rays, heating the
IGM and increasing the Jeans mass, have smoothed the gas on scales
(\ie, the ``filtering'' scale) larger than the size of the dark matter
filaments.

In \fig~\ref{fig:srei1}(left) we show the BH accretion rate (top
panel) and the baryon fraction in BHs, $\omega_{\rm BH}$ (bottom
panel), for the four simulations whose parameters are listed in
\tab~\ref{tab:1}.  
\begin{figure}
\centerline{\psfig{figure=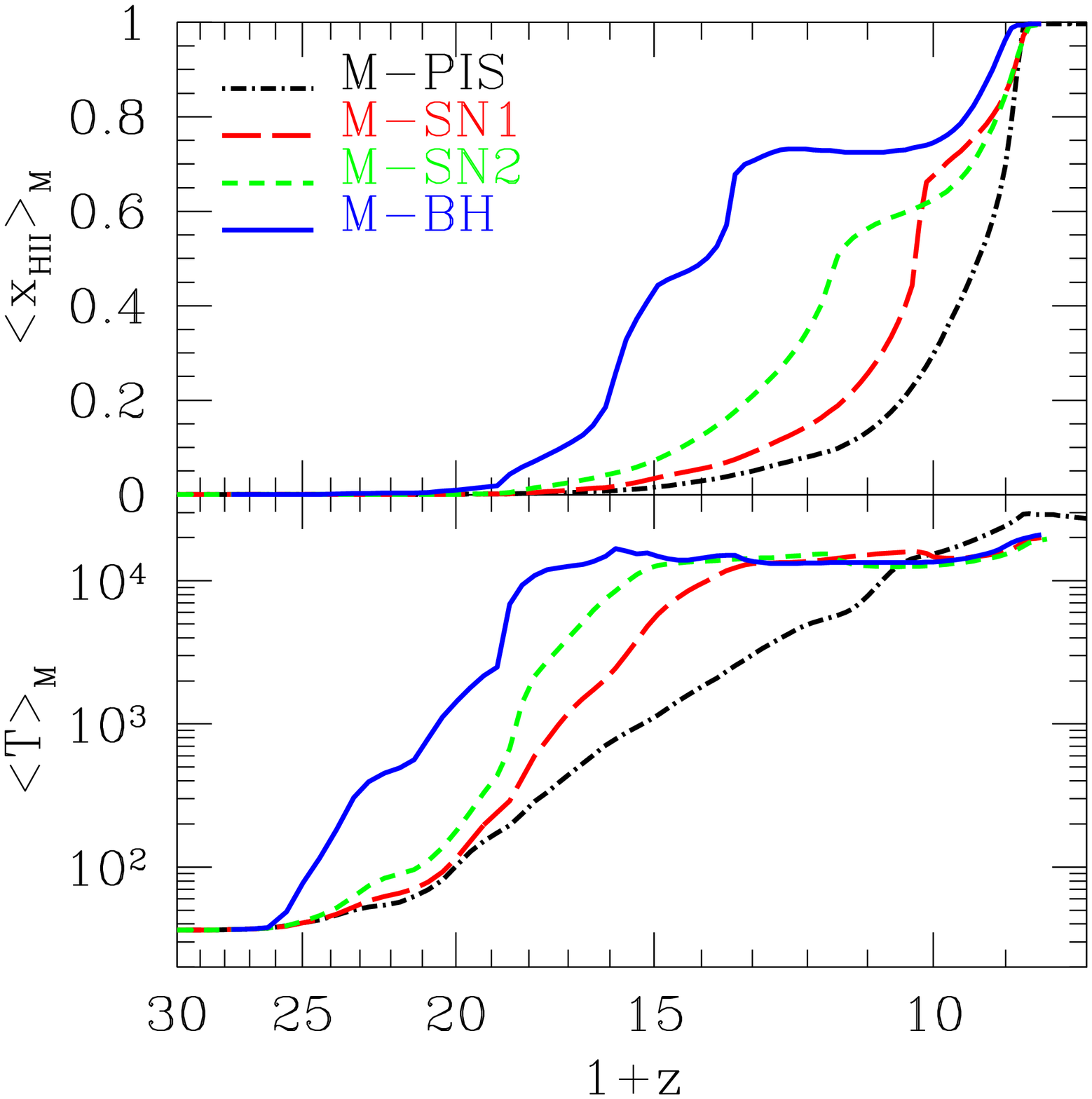,width=8.5cm}}
\caption{\label{fig:srei3} Hydrogen ionisation history (top panel) and
  thermal history (bottom panel) for the four models in
  \tab~\ref{tab:1}. In models of reionisation by \pop3 stars hydrogen
  has usually two distinct epoch of reionisation, the first at
  redshift $z \sim 17-10$ and the second at $z \sim 6-7$.  In the
  X-ray preionisation models hydrogen is partially reionised at early
  times but it is fully reionised only once at low redshift (\ie, by
  \popII stars at $z \sim 6-7$).  Note that the reheating of the gas
  always precedes the reionisation.  This characteristic is in common
  with all the reionisation scenarios.}
\end{figure}

In paper~IIa we have discussed the physical requirements needed to
produce such accretion histories. Here we do not discuss this matter
further but we note that, given the assumed SED of mini-quasars, it is
possible to determine an upper limit for the global accretion rate.
This limit is determined by the requirement that the early sources of
X-rays contribute to less than $10-20$\% of the observed X-ray
background in the $2-10$ keV bands. The sources that constitute the
bulk of the soft X-ray background has now been resolved and identified
(mainly Seyfert galaxies at $z \sim 1-2$). Recently \cite{deLuca:03}
have estimated that that about 20\% of the observed background may be
produced by a new population of faint X-ray sources, currently
undetected within the sensitivity limits of the deepest X-ray surveys.
It does not seem that the faint high redshift optically discoverable
sources are abundant enough to contribute much to the background
\citep{Hunt:03}. So there is some room for the population of sources
that we are postulating to exist at high redshift.  The solid line in
\fig~\ref{fig:srei1}(left) shows the value of the global accretion
rate as a function of redshift that would produce the 20\% of the
X-ray background at $z=0$ that it is still unaccounted for by lower
redshift AGNs.

In \fig~\ref{fig:srei1}(right) we show the star formation rate (top
panel) and the baryon fraction in stars, $\omega_{*}$ (bottom panel),
for the models in \tab~\ref{tab:1}.  Depending on the simulation
parameters, the thermal feedback on the IGM or internal feedbacks from
galactic winds due to photoevaporation and mechanical energy from SN
explosions, are dominant in reducing the global star formation rate in
small-mass galaxies \citep{OstrikerG:96, RicottiGSb:02}.
\begin{figure}
\centerline{\psfig{figure=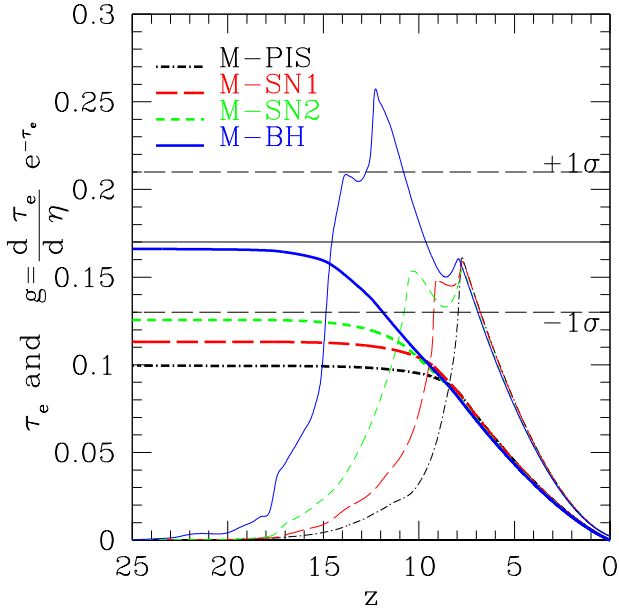,width=8.5cm}}
\caption{\label{fig:srei5} Thomson scattering optical depth, \taue,
  and visibility function, $g(z)$, as a function of redshift for the
  simulations in \tab~\ref{tab:1}. A simulation without X-ray
  preionisation with stellar reionisation by \popII stars at $z
  \approx 7$ would have \taue$=0.6$.}
\end{figure}

The results of the simulations confirm the results found using the
semianalytic models presented in paper~IIa.  The main signatures of
X-ray preionisation are illustrated in \fig~\ref{fig:srei3} that plot
the hydrogen ionisation history (top panel) and thermal history of the
IGM (bottom panel) as a function of redshift for the four simulations
in \tab~\ref{tab:1}. The early reheating of the IGM to a temperature
$T \approx 10^4$ K is reached when the hydrogen ionisation fraction
becomes larger than 10 \% (\cf, \figs~\ref{fig:srei3}). This happens
because the ionisation rate from energetic secondary photoelectrons
becomes inefficient when the fractional ionisation is larger than
10\%, consequently most of the energy of the X-ray photons is
deposited into heat. If the X-ray background is large enough, the soft
X-ray photons emitted from distant sources and redshifted into softer
UV photons, can reionise the voids above the 10\% fractional
ionisation.

The IGM optical depth to electron Thompson scattering, \taue, and the
visibility function, 
\begin{equation}
g(z)=\exp{(-\tau_e)}{d \tau_e \over d\eta},
\label{eq:vis}
\end{equation}
where $\eta$ is the conformal time, are plotted in
\fig~\ref{fig:srei5} for the simulations in \tab~\ref{tab:1}. We find
that, for the most extreme model (run number four in
\tab~\ref{tab:1}), X-ray preionisation can produce an optical depth to
Thompson scattering \taue$=0.17$, in agreement with WMAP measurement.
The shape of the visibility function determines the power spectrum at
large angular scales of the polarised CMB radiation (EE) and the
temperature-polarisation cross correlation (TE). In particular the EE
power spectrum can be used to distinguish between models with the same
\taue but different ionisation history because its shape is determined
by the redshift at which the visibility function has a peak
\citep{HolderHK:03}. As was found in paper~IIa the intermediate
preionisation scenario is the most efficient in producing a larger
optical depth to Thompson scattering because it is the earlier
preionisation scenario that requires the minimum number of ionising
photons per baryon (\ie, the recombinations are negligible).

\begin{figure}
\centerline{\psfig{figure=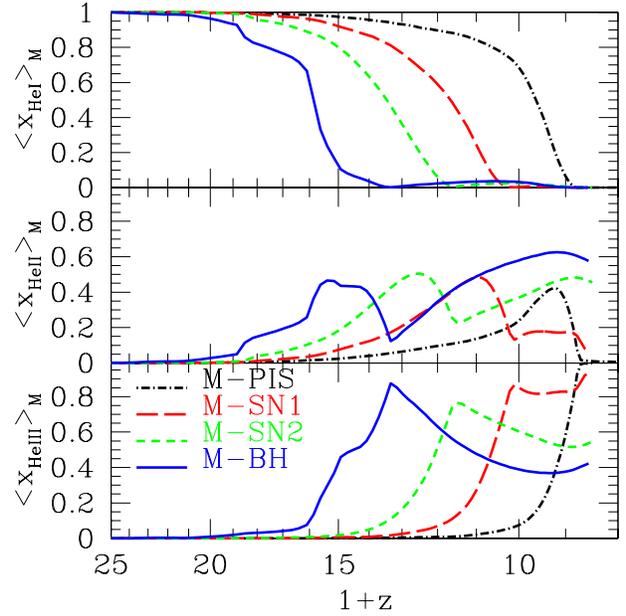,width=8.5cm}}
\caption{\label{fig:srei4} Helium ionisation history for the four 
  models in \tab~\ref{tab:1}. In most X-ray preionisation models \GII
  is reionised twice.  The helium is doubly ionised a first time at
  high redshift. Afterwards it partially recombines before redshift
  8-9. At redshifts $z \simlt 8-9$, due to the redshifted X-ray
  background, the ionisation fraction of \GIII starts increasing again
  and \GII become fully ionised a second time at $z \sim 3$ (see paper
  IIa for a discussion on this effect). This has also the effect of
  keeping the temperature of the IGM at about 10,000 K from redshift 6
  to 3.}
\end{figure}

Contrary to the double reionisation models proposed by \cite{Cen:03a},
in the X-ray preionisation models hydrogen is fully reionised only
once at low redshift (\eg, by \popII stars at $z \sim 6-7$). But in
most X-ray preionisation models \GII is reionised twice.  The helium
is doubly ionised a first time at high redshift. Afterward, in most
simulations, due to the fast decline in accretion rate onto seed BHs,
it partially recombines before redshift nine. But the rate of \GIII
recombination is slow because of the photoionisations from the
redshifted X-ray background photons.  This is shown in
\fig~\ref{fig:srei4}, where we plot the He ionisation history as a
function of redshift for the four simulations in \tab~\ref{tab:1}.  In
most simulations \GII becomes about fully ionised (60-80 \% fractional
ionisation) before \HI reionisation. It is possible to reionise \GII
before fully reionising \HI in the voids if it is the background that
dominates the ionisation rate (note that the background spectrum has
almost no UV photons).  Indeed the soft UV photons cannot ionise the
voids because they are absorbed locally while harder photons can still
ionise \HI but less efficiently than \GII.

At redshifts $z \simlt 9$, due to the redshifted X-ray background, the
ionisation fraction of \GIII starts increasing again and \GII becomes
raughtly fully ionised a second time at $z \sim 3$ (see paper IIa for
a discussion on this effect). The temperature of the IGM, due to \GII
ionisation, remains almost constant from redshift 6 to redshift 3 at
about 10,000 K (\cf, paper IIa), in agreement with observation of the
line widths of the \lya forest \citep{RicottiGS:00}.

\begin{figure}
\centerline{\psfig{figure=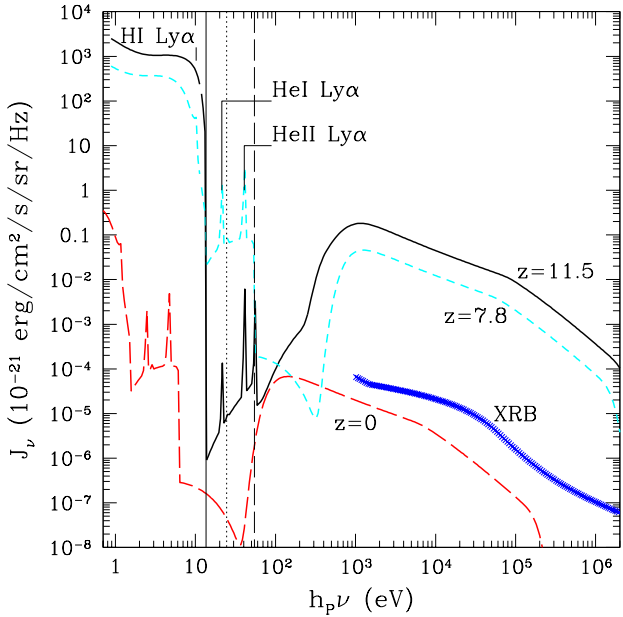,width=8.5cm}}
\caption{\label{fig:sback} Background radiation at $z=11.5, 7.8$ and
  the extrapolated value at $z=0$ for model M-BH. To guide the eye,
  the thin vertical lines show the energy of the Lyman continuum of
  \HI, \GI and \GII . The labels show the \lya lines of \HI, \GI and
  \GII.  The observed X-ray background radiation at $z=0$ is shown
  with a thick solid line. Even though early accretion on seed black
  holes produce only 10\% of the observed X-ray background (in the
  2-10 keV band), at redshift $z \sim 3$ they may have been
  predominant, indeed most of the X-ray background is produced by
  Seyfert galaxies at $z \sim 1$.}
\end{figure}

In \fig~\ref{fig:sback} is shown the background radiation at $z=11.5,
7.8$ and the extrapolated value at $z=0$ for model M-BH. The observed
X-ray background radiation at $z=0$ is shown with a thick solid line.
Even though early accretion on seed black holes produces only 10\% of
the observed X-ray background (in the 2-10 keV band), at redshift $z
\sim 3$ they may have been predominant, indeed most of the X-ray
background is produced by Seyfert galaxies at $z \sim 1$.
Our analysis of the model constraints posed by the observed unresolved
component of the X-ray background (presented in paper~IIa and also
shown here in \fig~\ref{fig:sback}) are consistent with the findings
of \citep{Dijkstra:04}. They also do not rule out a high redshift
mini-quasar population that could partially reionise the IGM to 50\%
ionisation fraction. In particular, in the case that the quasar
spectrum has only a soft X-ray component (\eg, produced by the hot
multicolour accretion disk) and no hard X-ray emission, the X-ray
background does not pose any constraint on the number density and
luminosity of a putative high redshift population of X-ray sources.
Sources with this SED could preionise the IGM very efficiently.

\begin{figure*}
\centerline{\psfig{figure=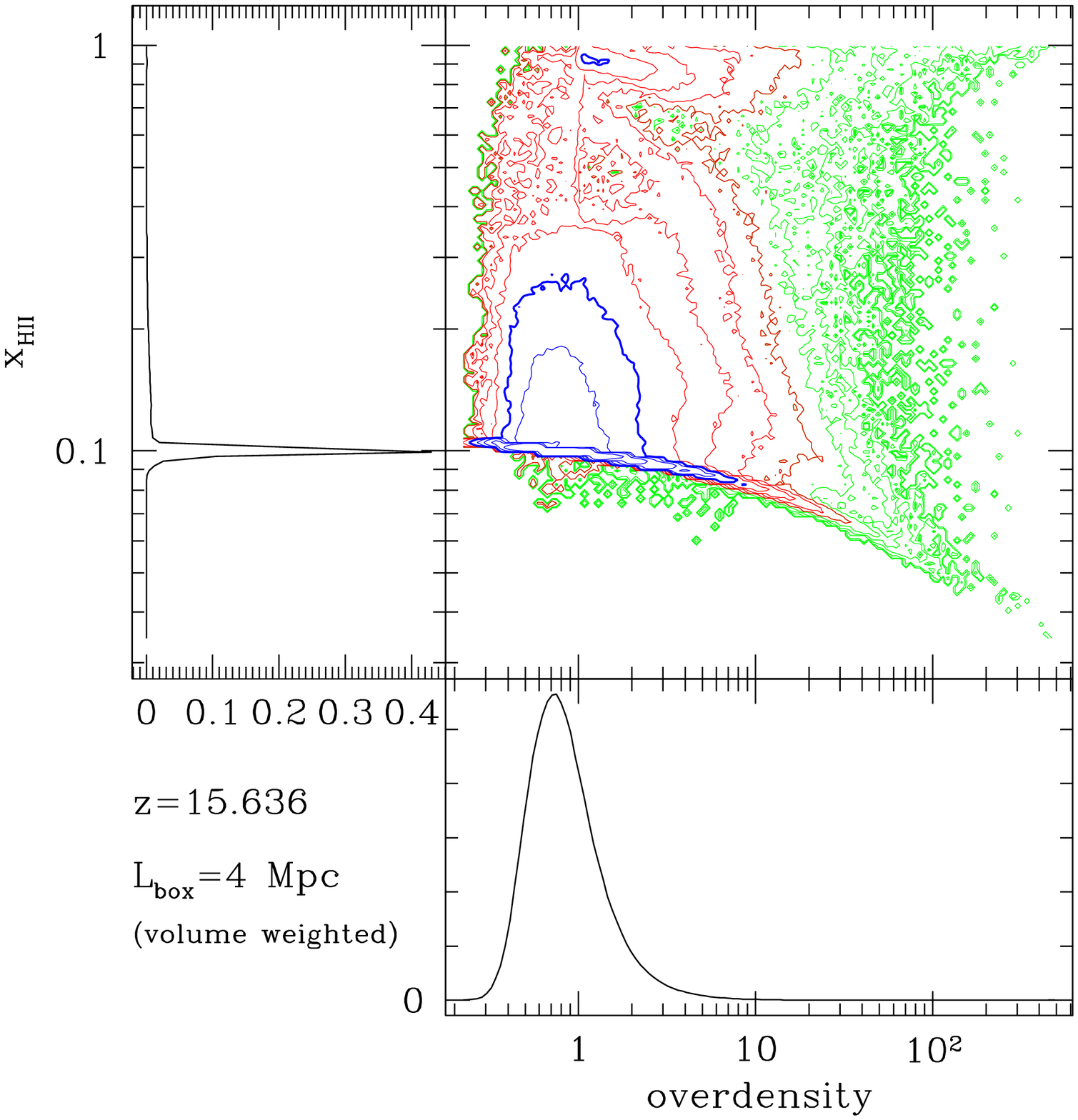,width=8.5cm}
\psfig{figure=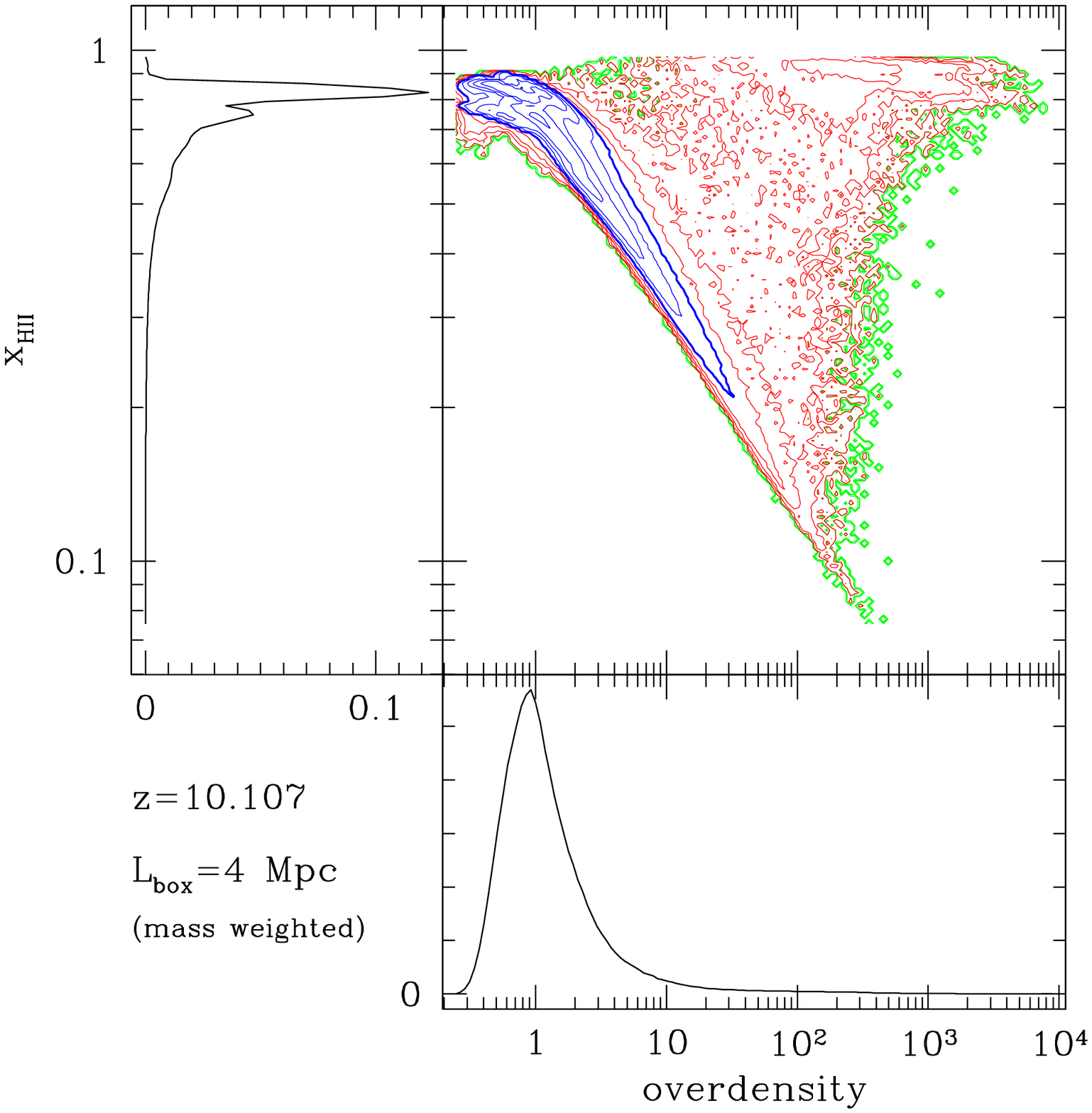,width=8.5cm}}
\caption{\label{fig:hiiov} Mass-weighted distribution of the mean
  hydrogen ionisation fraction versus the overdensity for the M-BH
  simulation at $z \simeq 15$ (left) and $z \simeq 10$ (right). The
  stellar sources did not reionise the IGM at this redshift but the
  underdense regions are already almost fully reionised. UV photons
  from the redshifted X-ray background ionise the IGM, preferentially
  the low density regions.  The voids remain ionised because the
  recombination time is longer than the Hubble time.}
\end{figure*}

The efficiency of secondary ionisation from energetic photo-electrons
is large when the ionisation fraction of the IGM is less than 10\%.
This is illustrated in \fig~\ref{fig:hiiov} that shows the
mass-weighted distribution of the mean hydrogen ionisation,
$x_{\HII}$, versus the overdensity for the M-BH simulation at redshift
$z \simeq 15$ (figure on the left) and $10$ (figure on the right).  At
redshift $z \sim 15$ the hydrogen in most of the IGM volume has a
fractional ionisation of 10\% or less as expected.  Interestingly, at
$z \simeq 10$, even if reionisation by stellar sources is not
complete, the underdense regions are already almost fully ionised.
This is due to the intense background in the UV bands arising from the
redshifted X-rays emitted by distant sources that ionises the IGM,
preferentially in the low density regions that occupy most of the
volume. The ionisation fraction in the voids remains large because the
recombination time is longer than the Hubble time in these underdense
regions.  Because the atomic hydrogen in the low density intergalactic
medium is almost fully ionised before stellar reionisation at redshift
$z \sim 6-7$, it follows that stellar reionisation is characterised by
an almost instantaneous ``overlap phase'' of \HII regions.

In the remaining paragraphs of this section we describe the results of
each simulation in greater detail.

In the simulation M-PIS (dot dashed line), the IMF is top-heavy,
\fesc$=1$ and the energy input from SN explosions is 7 times larger
than for a Salpeter IMF. This case is one in which pair-instability
SNe are important in polluting the IGM and in producing strong
mechanical feedback in the ISM. The star formation is strongly
suppressed by the SN explosions. At $z \approx 7$ the SFR is $5 \times
10^{-3}$ M$_\odot$ Mpc$^{-3}$s$^{-1}$, 20 times smaller than the
observed value at $z \sim 4-6$. The box size of 1 \Mpc is too small to
achieve convergence at $z \simlt 12$ (this is demonstrated more
quantitatively in paper~I), therefore at those redshifts
the star formation is probably underestimated. But the result shows
that star formation is strongly suppressed in galaxies less massive
than $10^9$ M$_\odot$ for this choice of the IMF. Metal pollution from
SN explosions may suppress the formation of \pop3 stars before $z=10$,
when we stop our simulation. This would reduce the UV emissivity,
$\epsilon_{\rm UV}$, and the SN energy output, $F_{\rm IMF}$. Here, we
did not take into account this effect.  In this simulation the assumed
efficiency of X-ray emission, $\epsilon_{\rm qso}=2 \times 10^{-3}$,
is too small to have a substantial effect on the ionisation and
re-heating of the IGM.  Reionisation of \HI and \GII happens at $z
\approx 8$ and the gas is heated to $T \approx 3 \times 10^4$ K.

In the simulation M-SN1 (dashed line), we assume a Salpeter IMF and
\fesc$=1$. Here the global SFR is about 10 times larger than in the
M-PIS simulation, because of the less violent feedback from SN
explosions. The increased SFR produces a larger X-ray emission that at
$z \simeq 12$ re-heats the IGM to $T \simeq 10^4$ K and partially
ionises the IGM \xe$\simeq 10$\%. The X-ray emission continues to
increase until redshift $z =10$ as a result of the increasing star
formation and the electron fraction reaches \xe$\sim 50$\%. At $z \sim
9$ \GII is reionised. In this model after $z_{\rm off}=10$ we assume
zero X-ray emissivity ($\epsilon_{\rm qso}=0$). The \GII starts
recombining but the ionisation fraction continues to increase, but
more slowly.  This effect is due to UV photons arising from redshifted
X-rays of the background radiation. For this model \taue$\sim 0.11$ and
the visibility function peaks between redshifts $6<z<9$.

The simulation M-SN2 (short dashed line) is the same as M-SN1
but has \fesc$=10$\% and the effect of SN explosions is $10$ times
smaller. This assumption is justified by the fact that
zero-metallicity stars may collapse directly into BHs without
exploding as SN. Another possibility is that the energy of SN
explosions in zero-metallicity stars is smaller than the canonical
value $E=10^{51}$ ergs \citep{UmedaN:03}. The results for this
simulation are analogous to M-SN1 but by virtue of the larger SFR
(by three times), partial X-ray ionisation and re-heating starts
earlier (at $z \sim 14$) and \taue$\sim 0.13$.
 
Finally, simulation M-BH (solid line) is the same as M-SN2 but
has a larger X-ray emissivity, $\epsilon_{\rm qso}$. Contrary to
previous models where we have assumed a constant value of
$\epsilon_{\rm qso}$, here the emissivity is a fit to the
semianalytic intermediate preionisation model (model M3 in paper~IIa).
In this simulation the thermal feedback produced by the X-ray
background on the SFR is evident. At $z>15$ the global SFR is reduced
by about one order of magnitude. It starts increasing again and
reaches the same magnitude as in the M-SN2 simulation at $z =12$
when more massive galaxies start to be numerous. The X-ray sources are
in the most massive galaxies in which the star formation rate (and the
accretion rate) is not suppressed by feedback processes. This can be
seen also in the bottom panels of \fig~\ref{fig:slice}.  The X-ray
background re-heats and partially ionises the IGM starting at redshift
$z \sim 20$. At $z \sim 13$ \GII is almost completely reionised and
\xe$\sim 0.7$.  Afterwards, \GII slowly recombines but \HI remains
partially ionised until $z=7$, when reionisation by stellar sources is
completed.
\begin{table}
\caption{Results of the hydrodynamic simulations with radiative
  transfer.\label{tab:2}}
{\footnotesize
\begin{tabular*}{8.3 cm}[]{l|cccccl}
{RUN} & \taue & $z_g^{max}$ & $y$-par &
$\omega_{\rm BH}$ &  $f_{\rm XRB}$ & $\Gamma(\GII) t_{\rm
  H}$\\
 &  &  & $\times 10^{-7}$ & $\times 10^{-5}$ & (\%) & (\%)\\
\hline
M-PIS & 0.10 & 8 & $3.0$ & 0.1  & 19 (12) & $>200$\\
M-SN1 & 0.11 & 9 & $3.1$ & 0.2  & 11 (6) &  $60$\\
M-SN2 & 0.13 & 10.5 & $3.3$ & 0.3  & 9 (5) & $25$\\
M-BH  & 0.17 & 12 & $4.4$ & 2.0  & 17 (8) &  $20$\\
\end{tabular*}}
\\

Meaning of the values in each column: \taue is the Thomson scattering
optical depth; $z_g^{max}$ is the
high-redshift maximum of the visibility function $g(z)$ (it also
roughly coincides with
the redshift of early reionisation of \GII; $y$ is the
Compton distortion parameter; $\omega_{\rm BH}$ is mass fraction in BH
in units of the baryon cosmic
density at redshift $z=8$ (to translate $\omega_{\rm BH}$ to
cosmic BH mass density, $\rho_{\rm BH}$, multiply by $~5.5 \times
10^9$ M$_\odot$ Mpc$^-3$); 
$f_{\rm XRB}$ is the fraction of the X-ray background at 50-100
keV (and 2-10 keV, in parenthesis) due to early black holes; 
$\Gamma(\GII) t_{\rm H} \sim (t_H/n_{\GII}) dn_{\GII} /dt$ is the 
fractional rate of \GII photoionisation per Hubble time per helium 
atom at redshift $z \sim 2-3$ due to the redshifted X-ray background.
\end{table}

In \tab~\ref{tab:2} we summarise the values of some relevant
quantities computed from the simulations in \tab~\ref{tab:1}. The
meaning of the quantities shown in each column of the table are given
in the footnote.

\subsection{Dependence on the assumed mini-quasars
  spectra}\label{ssec:sed}

We have seen that in our models an important contribution to \HI and
\GII reionisation is produced by redshifted soft X-rays.  Therefore if
the quasar spectrum is strongly absorbed we expect that the importance of
X-rays for the \HI preionisation and \GII double reionisation may be
reduced.

Our template spectrum has a cutoff at photon energies of a few 100~eV
and therefore differs from \cite{SazonovO:04} absorbed spectrum that
does not have emission below few keV.  We have assumed that the
mini-quasar spectrum is absorbed by a column density of H and He with
neutral hydrogen column density $N_{\rm ab} \simeq 10^{19}$ cm\mm (see
\S~\ref{ssec:code}), substantially smaller than in \cite{SazonovO:04}.
There is no strong theoretical argument in favour of high redshift
quasars being strongly absorbed and having no soft X-ray emission.  On
the contrary, high redshift galaxies have gas components with radii of
a few 100~pc and mean gas densities of $10-100$ cm\mmm. The hydrogen
column density is therefore only $N_{\rm ab} \sim (x_\HI/0.001)
10^{19}$ cm\mmm, where $x_\HI$ is the hydrogen neutral fraction. Most
important, the metallicity is expected to be subsolar ($Z \sim 0.01$
Z$_\odot$) and in highly absorbed quasars most of the X-ray absorption
is produced by metal lines. The \cite{SazonovO:04} template is
therefore more appropriate for quasars at redshifts $z \simlt
5$, hosted by quite massive galaxies.
\begin{figure}
\centerline{\psfig{figure=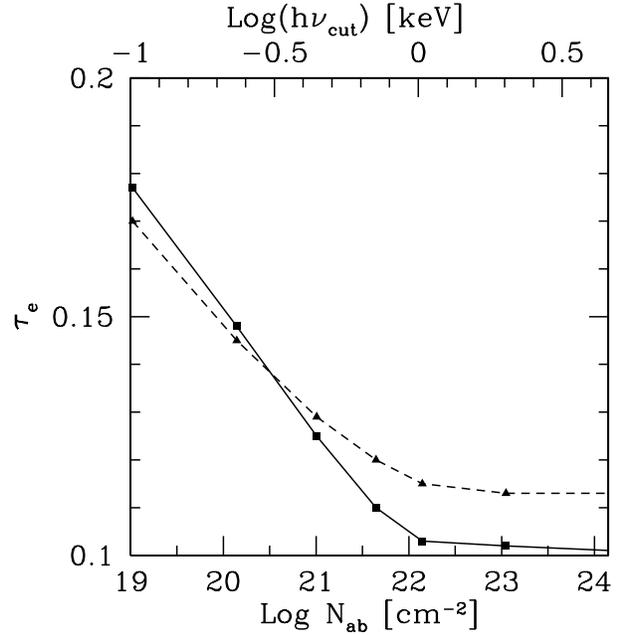,width=8.5cm}}
\caption{\label{fig:Na} Dependence of the optical depth to Thompson
  scattering of the IGM, \taue, on the average absorbed mini-quasar
  spectrum. We show \taue as a function of the column density of
  neutral hydrogen absorbing the quasar spectrum or, equivalently, the
  cutoff energy of UV and soft X-ray photons in the spectrum (shown on
  the top axis) for model M4 in paper~IIa assuming \pop3 star SED
  (dashed line) or \popII star SED (solid line). The results (shown by
  the points) are calculated using a semianalytic model (see
  paper~IIa) that is equivalent to run M-BH in \tab~\ref{tab:1}.}
\end{figure}

More quantitatively, our assumption can be justified by the following
arguments.  In our simulations the spatial resolution is about $r=50$
pc therefore, for the assumed column density $N_{\rm ab}=n_\HI r$, we
have $n_\HI =n x_\HI \sim 0.1$ cm\mmm, where $n$ is the gas density and
$x_\HI$ is the hydrogen neutral fraction.  For a gas in
photoionisation equilibrium we have
\begin{equation}
x_\HI= (3 \times 10^{-4}~{\rm cm}^{-3}) \left({n \over 10~{\rm
      cm}^{-3}}\right) \left({R \over 50~{\rm pc}}\right)^2 \left({S_0
      \over 10^{51}~{\rm s}^{-1}}\right)^{-1} ,
\end{equation}
where $S_0$ is the number of ionising photons emitted per second and
$R$ is the distance from the source. For example, a $10^4$ M$_\odot$
black hole accreting at the Eddington limit would emit about $S_0 \sim
10^{50}-10^{51}$ s\m EUV photons. Given our choice of the absorbed
spectrum we infer a density profile inside our resolution element $n =
(60~{\rm cm}^{-3}) (50~{\rm pc}/R) (S_0/10^{51}~{\rm s}^{-1})^{1/2}$,
that seems a quite reasonable assumption.

Nevertheless, it is interesting to study how our results would change
if the spectrum has a cutoff at energies $h\nu_{\rm cut} > 100$ eV.
We have run a set of semianalytic models assuming that the X-ray
sources are absorbed by larger \HI column densities ranging from
$N_{\rm ab} =10^{19}$ cm\mm to $10^{25}$ cm$^-2$ (approximately we
have $h\nu_{\rm cut} \sim 100 eV (N_{\rm ab}/10^{19}~{\rm
  cm}^{-2})^{1/3}$). We found that \taue decreases linearly with
increasing $\log N_{\rm ab}$ and reaches an asymptotic minimum value
of \taue$ \sim 0.1$ for $N > 10^{22}$ cm\mm.  In \fig~\ref{fig:Na} we
show the dependence of the optical depth to Thompson scattering of the
IGM, \taue, on the average absorbed mini-quasar spectrum. We can fit
the result shown by the solid line with the function
\begin{equation}
\tau_e = {\rm Min}[0.177 - 0.026(\log N_{\rm ab}-19), 0.1].
\end{equation}

The reionisation of \GII at $z \sim 13$ shown in \fig~\ref{fig:srei4},
is also produced by soft X-ray photons emitted by mini-quasars.
Increasing the absorbing column density of the quasar spectrum
produces a high redshift reionisation only in the low density regions
and a partial ionisation in the overdense regions of the IGM.  When
the column density is $N_{\rm ab}> 10^{21}-10^{22}$ cm\mm, the
high-redshift reionisation does not happen at all. Similarly, the
importance of redshifted X-rays for the low redshift reionisation
decreases and is slightly delayed to redshifts $z \sim 2-3$ when the
absorbing column density increases. Nevertheless, even for large
column densities ($N_{\rm ab} \sim 10^{24}$ cm$^{-2}$), redshifted
X-rays are still able to reionise \GII in underdense regions at $z
\sim 2$ and partially ionise the overdense regions. Though, if we
assume a \pop3 SED for the stars and \HI reionisation at $z \sim 6.5$,
then \GII reionisation is dominated by the stars and happens earlier,
at $z \sim 3-4$. Note that we do not include the contribution from
observed AGNs at redshifts $z \simlt 5$. Perhaps these AGNs are
sufficient to produce \GII reionisation and IGM reheating at $z \sim
3$ without the need for additional contributions.

Finally, as mentioned earlier, it is plausible that the spectra of
high redshift X-ray sources have only a soft X-ray component produced
by thermal emission from a multicolour accretion disk and no hard
X-ray emission.  Accreting intermediate mass black holes, because the
disk is hotter than in supermassive black holes, emit in the FUV and
soft X-rays.  This would be the most favourable scenario for achieving
large values of \taue since soft X-rays efficiently pre-ionise the IGM
but high redshift sources, even if very numerous, would be invisible
in X-ray and optical deep fields and would not contribute to the
observed X-ray background.
 
\section{Number counts of point sources in the X-ray bands}\label{sec:lum}

We have shown that the high redshift X-ray sources (that we postulate
to produced the observed \taue of the IGM) produce less than 20\% of
the observed X-ray background. In this section we calculate their
contribution to the luminosity function of X-ray point sources.
Available Chandra deep field observations can already constrain the
most extreme models \citep[paper~IIa,][]{Dijkstra:04} and the future
observations with Constellation-X \citep{White:99} and XEUS will be a
factor $\sim10$ times fainter than the current deep survey limit. 

It is important to notice that in these calculations we assume a
spectral energy distribution of the sources that has a non-negligible
high energy power-law component (\cf, \fig~4 in paper~IIa), in
agreement with observations of QSOs and ULX in the local universe. But
if the source spectra are dominated by a multicolour disk thermal
component their contribution to both the X-ray background in the 2-50
keV bands and the faint source counts would be negligible. Instead the
full ionisation of atomic hydrogen in the low density IGM before
redshift $z \sim 7$, the \GII reionisation at $z \sim 3$ and reheating
of the low-redshift IGM is produced by the redshifted X-ray background
independently of the assumed spectra of these early sources.

In our models we do not make assumptions about the mass function of
BHs in each galaxy, but only on their total mass and accretion
efficiency. Two scenarios are therefore plausible for the X-ray
emission. If seed BHs quickly merge into a massive BH at the centre of
galaxies through dynamical friction, the X-ray sources will look like
small AGNs (mini-quasars).  If the seed BHs do not merge efficiently,
in each galaxy we could have several off-centre X-ray sources produced
by accreting intermediate mass BHs orbiting the galaxy. These sources
would be similar to ULXs observed in nearby galaxies, but more
numerous (tens or hundreds per galaxy) or more luminous. This new type
of X-ray sources could be common in the early universe if primordial
galaxies are compact and gas rich. This is expected in young galaxies
with masses $M_{\rm dm} \simgt 10^8$ M$_\odot$ in which stellar winds
did not evaporate all their gas \citep{Whalen:03, RicottiGS:04} and
star formation did not have time to consume most of the gas. Moreover,
in the preionisation models, the number density of intermediate mass
BHs was larger in primordial galaxies than in today's galaxies as was
shown in \fig~3 of paper~IIa.
\begin{figure}
\centerline{\psfig{figure=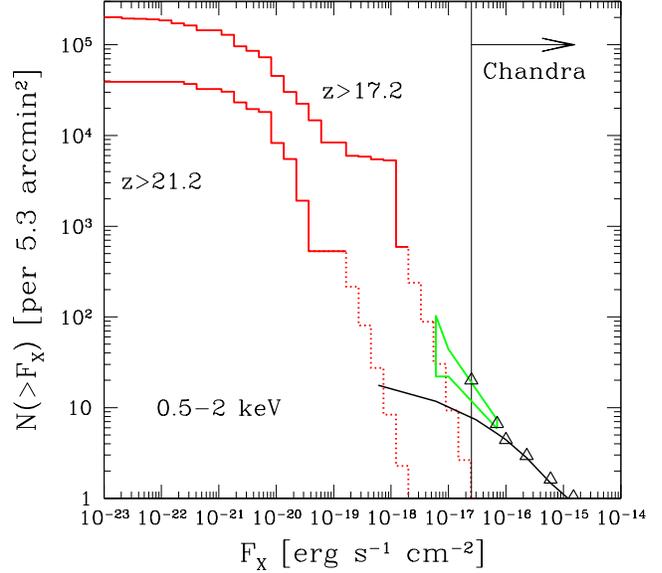,width=8.5cm}}
\caption{\label{fig:lumin} Number counts of X-ray sources in the 0.5-2
  keV band for the M-BH run (solid histogram). The brighter tail of
  the luminosity function is produced by the most massive galaxies at
  redshift $z \sim 17$ hosting either a supermassive BH of about
  $10^5$ M$_\odot$ accreting at near the Eddington rate or several
  intermediate-mass BHs with the same total mass. The dashed portion
  of the histogram is an extrapolation based on the Press-Schechter
  formalism. The thin vertical line shows the sensitivity limit of
  $F_{\rm X}=2.5 \times 10^{-17}$ erg s\m cm\mm, of the 2 Ms Chandra
  deep field in the 0.5-2 keV band, the triangles show the number
  counts of point sources and the solid line a model prediction for
  the expected counts from AGNs.  Finally the closed polygon shows the
  expected steepening of the counts based on the statistics of pixel
  fluctuation from unresolved point sources. The observed steepening
  of the number counts slope at fluxes $F_{\rm X} \simlt 10^{-16}$ is
  consistent with the predictions of our simulation but it is also the
  predicted contribution of starburst galaxies. The high-z sources,
  contrary to starburst galaxies, do not have optical counterparts but
  could be detectable in the infrared if they are not obscured locally
  by dust.}
\end{figure}

The luminosity of accreting BHs in each host galaxy is, by assumption,
proportional to the star formation rate in that galaxy, that is, to
the ability of the gas in that galaxy to cool efficiently and be
accreted toward the centre. Due to the early reheating by the X-ray
background and feedback processes (\ie, galactic winds produced by
photoionisation and SN explosions) the star formation in the smaller
mass galaxies is strongly suppressed and so is the accretion rate on
intermediate mass BHs. Moderately larger galaxies are able to retain
their gas but they cool so inefficiently that few or no stars at all
are formed. When seed BHs are accreted on these galaxies they find a
large reservoir of gas and they can start accreting efficiently.

The X-ray luminosity function, shown in \fig~\ref{fig:lumin}, is
dominated by the most massive galaxies that are less affected by
feedback effects and can retain gas more effectively.  This means that
the most rare and massive galaxies at $z \sim 15$ are expected to be
the most luminous X-ray sources. The volume of our simulation (1
\Mpc)$^3$ is large enough to be a representative region of the
universe at $z \sim 15$ but it is too small to contain any of these
rarer objects. We find that the luminosity of the typical X-ray
sources at $z \sim 15$ is two orders of magnitude fainter than Chandra
deep field 3$\sigma$ detection limit for $1.945 \times 10^6$ seconds
integration time \citep{Alexander:03} (limiting flux of $2.5 \times
10^{-17}(1.4 \times 10^{-16})$ erg s\m cm\mm sr\m in the $0.5-2(2-8)$
keV band). But we expect a few thousands of these sources per
arcmin$^{2}$.  We then use the Press-Schechter formalism to
extrapolate the number counts found in our simulation to larger
volumes. The extrapolated high luminosity tail of the number counts is
shown in \fig~\ref{fig:lumin} with a dotted line. The plot shows that
it is plausible to find a few sources per arcmin$^{2}$ above the
detection limit (shown by the vertical line) of the Chandra X-ray deep
field (unless the high tail of the luminosity function has a sharper
cutoff than in the semianalytic predictions).  The triangles show the
number counts of point sources of the 2 Ms Chandra deep field in the
0.5-2 keV band and the closed polygon shows the expected slope of the
luminosity function based on the statistics of pixel fluctuation from
unresolved point sources \cite{Miyaji:02}. The solid line that fits
the number counts is a model prediction for the expected counts from
AGNs \citep[\eg,][]{Comastri:95}.  We see that the observed steepening
of the number counts slope at fluxes $F_{\rm X} \simlt 10^{-16}$ is
consistent with the predictions of our simulation but it is also
consistent with the predicted X-ray contribution by starburst galaxies
\citep[\eg,][]{Ptak:01,Ranalli:03}.  In principle it is possible to
separate the contribution of these two populations since the high-z
sources, contrary to starburst galaxies, do not have optical
counterparts but may be detectable in the infrared if they are not
obscured locally by dust.  The planed spatial resolution of
Constellation-X and XEUS of about 5 arcsec is not sufficient to avoid
source confusion (Mushotzky, private communication). The cross
correlation of the unresolved X-ray background fluctuation with the
optical number counts may be used to separate the contribution of the
high redshift population from the starburst population.

This prediction has exciting consequences for the direct observability
of X-ray sources at redshifts as high as $z \sim 15$. Interestingly,
\cite{Koekemoer:03} describe a possible new class of X-ray sources
that have robust detections in ultra-deep Chandra data, yet have no
optical counterpart in deep multi-band GOODS Hubble Space Telescope
(HST) ACS images.  Their ratios of X-ray to optical fluxes are at
least an order of magnitude above those generally found for other AGN,
even those that are harboured by reddened hosts. The authors conclude
that if these sources lie above redshifts 6-7, such that even their
\lya emission is redshifted out of the bandpass of the ACS z(850)
filter, then their optical and X-ray fluxes can be accounted for. They
find seven of these sources in the field, in agreement with our
estimates for the M-BH run. In our model, if the source emission
extends to the rest frame visible/UV bands (\ie, if they are not
heavily reddened by dust), they should be detectable in the infrared
above 2-3 micron. Otherwise they should be visible only in X-ray bands
and in the FIR at about 100 $\mu$m.

The postulated existence of this mini-quasar population increases the
probability of finding a few radio loud sources at $z \sim 17$, that
could be successfully used to study the redshifted 21 cm absorption
lines produced by the hyperfine transition of neutral hydrogen in the
IGM before reionisation \citep{Carilli:02}. Next we discuss further
the radio signal at $21(z+1)$ cm.
\begin{figure}
\centerline{\psfig{figure=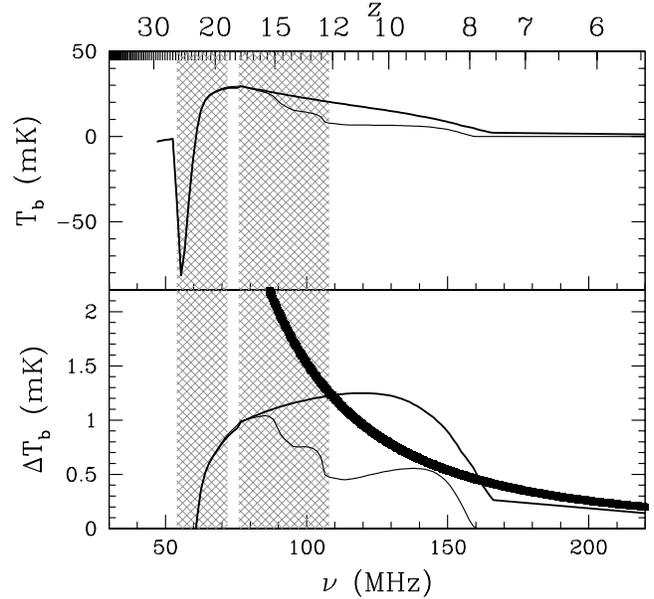,width=8.5cm}}
\caption{\label{figFromNick} Top panel: the mean redshifted 21 cm signal as
  a function of frequency and redshift for the simulation run M-BH
  (thin lines) and our fiducial semi-analytical model (thick lines).
  The two hatched zones show the locations of TV and FM radio bands
  within which is likely that no observations will be possible due to
  man-made interference.  Bottom panel: the fluctuations in the
  redshifted 21 cm signal in a beam of $10^\prime$ and $1{\rm MHz}$
  bandwidth.  The solid black band is the sensitivity limit of a
  filled dish aperture telescope for 1000 hours of integration. }
\end{figure}

\section{Redshifted 21 cm line in emission and absorption}\label{sec:21cm}

One possible observational test for our model is the detection of
redshifted 21 cm radiation from high redshift \citep*{Madau:97,
  Tozzi:00}. \fig~\ref{figFromNick} shows the predicted mean and rms
signal for our representative simulation and the best-fit fiducial
model. The only instrument capable of detecting the redshifted 21 cm
signal is the projected Low Frequency Array (LOFAR).  The solid black
band shows the expected LOFAR sensitivity for observations with the
core (2 km in diameter) for 1000 hour integration time.

In calculating the expected 21 cm signal, we included three physical
effects that influence the level population for the hyper-fine
transition in hydrogen atom: pumping by \lya photons and
collisions with electrons and neutral atoms \citep{Tozzi:00}.  The
former is the dominant effect at lower redshifts ($z<20$), while the
latter is important at higher redshifts. We then simulated a number of
lines of sight through the computational box and averaged over them to
properly include the effect of velocity focusing: because most of the
emission comes from the high density regions, which are collapsing,
velocity focusing increases the effect by a small but not negligible
factor.

Fluctuations in the expected 21 cm signal, however, come from large
spatial scales (larger than about 10 Mpc). The density fluctuations in
this regime are linear, and can be computed analytically (Gnedin \&
Shaver 2004).

As one can see from \fig~\ref{figFromNick}, the mean signal is well
within the sensitivity of existing or proposed low frequency radio
telescopes (such as Arecibo, LOFAR, or SKA).  However, the main
challenge to observing the mean signal is not the sensitivity, but
rather the foreground contamination from the Galaxy
\citep[\cf,][]{OhM:03}.  The smoothly variable emission signal will
most likely be unobservable, but in our model a remarkable opportunity
arises: a sharp absorption feature at $z\sim 25$ falls just outside
the broadcast TV band (that starts at 54 MHz, which corresponds to
$z=25$). This feature will appear as an about 100 mK ``absorption
line'' in the spectrum of the Galactic foreground, and may be
observable with future radio telescopes.  The specific location of
this feature is, of course, model dependent - a slight variation of
cosmological parameters well within WMAP errors can either move it to
higher redshifts, or hide it entirely behind the broadcast TV band.
Thus in the later case, observing it will probably be impossible, but
if indeed it falls beyond the broadcast TV band, it will clearly
indicate the ``first light'' in the Universe - or, at least, first
\lya photons.

The observational situation with fluctuations is, in some sense,
reversed with respect to the mean signal. While it will be
significantly easier to separate fluctuations in redshifted 21 cm
emission from the Galactic foreground, the early emission of
\lya photons relative to the growth of structure makes the
fluctuation signal harder to observe than in more conventional
reionisation models \citep{GnedinS:03}.  In addition, at the predicted
level of the cosmological signal, systematic errors such as beam
leakage also become important. It, therefore, appears that unless the
absorption feature at 54 MHz can be detected, existing and projected
radio telescopes (including LOFAR and SKA) will not be able to detect
cosmological fluctuations in the redshifted 21 cm signal in emission
(but there still remain the possibility to detect the 21 cm forest in
absorption against high redshift radio sources).
\begin{figure}
\centerline{\psfig{figure=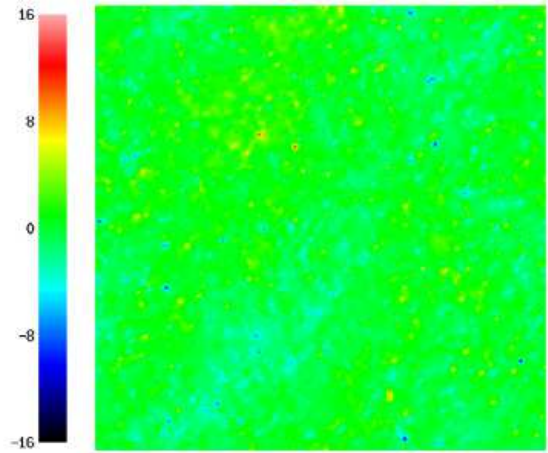,width=7.5cm}}
\caption{\label{figFromNickCMBmap} Secondary temperature anisotropies
  (in $\mu$K) in a $0.55 \times 0.55$ arcmin$^2$ patch of the sky for
  the run M-BH. The map shows strongly non-linear features produced
  by gas in minihaloes.}
\end{figure}

\section{Secondary CMB anisotropies}\label{sec:cmb}

\begin{figure}
\centerline{\psfig{figure=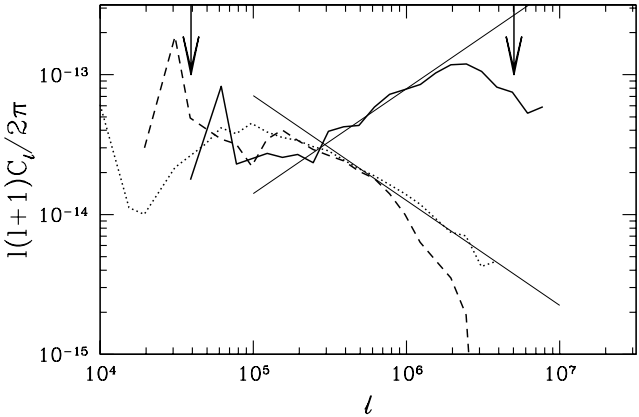,width=8.5cm}}
\caption{\label{figFromNickCMB}The power spectrum of the CMB anisotropies
  for the run {\bf M-BH} (solid thick line). For comparison, the
  dashed and dotted lines show the power spectrum for two models
  without early X-ray sources: the dashed lines is the simulation {\bf
    128L2noSN} with box size of 2 \Mpc (from Ricotti \& Ostriker 2003)
  and the dotted line is a simulation with box size of 4 \Mpc (from
  Gnedin \& Jaffe 2000). Two thin straight lines mark +0.75 and -0.75
  slopes respectively. Two vertical arrows show the fundamental and
  the Nyquist frequencies of the simulated sky for the run {\bf
    M-BH}.}
\end{figure}

Another possible observational signature of the early episode of X-ray
emission is secondary CMB anisotropies. We used \cite{GnedinJ:01}
method to compute the spectrum of secondary anisotropies on small
angular scales (arcmin$^2$ or $l>10^4$), where the anisotropies from
the first episode of structure formation dominate other contributions.
Maps of the temperature anisotropies for the X-ray preionisation
simulation (run M-BH, left panel) and \pop3 stellar reionisation
simulation (run 128L2VM in paper~I, right panel) are shown in
\fig~\ref{figFromNickCMB}.

\fig~\ref{figFromNickCMB} shows the angular spectrum for runs M-BH and
128L2VM together with the typical spectrum from \cite{GnedinJ:01}.  A
remarkable feature can be observed: while in a model without early
episode of X-ray emission $C_ll(l+1)$ falls off at small angular
scales approximately as $l^{-3/4}$, in our model the anisotropies
instead grow roughly as $l^{3/4}$.  The temperature anisotropies are
of the order of $\Delta T \sim 16 \mu$K at $\sim 1$ arcsec scales. 
This opens up a unique opportunity to test our models against those
without early episode of X-ray emission, although, observations on
these scales are in the future and may be extremely difficult to
separate the signal from the foreground sources contamination
\citep[but see,][]{Fomalont:93,Church:97}. 



The non-linear Ostriker-Vishniac effect \citep{OstrikerV:86} describes
anisotropies generated in a universe in which the ionisation fraction
is homogeneous in space. In our simulations the ionisation fraction is
not homogeneous in space and the total power spectrum of the
anisotropies includes a contribution from the ``patchy reionisation''.
In the X-ray preionisation model the non-linear OV component of the
anisotropy power spectrum (calculated assuming the mean ionisation
fraction) is actually slightly larger than the total signal, which
means that the ``patchy reionisation'' component is mildly
anti-correlated with the OV component. In other words, the higher
density regions are less ionised than the low density regions - this
is, obviously, a signature of X-ray ionisation.  Such a separation is
purely artificial and unphysical, but illustrates the main physical
difference with the UV-ionisation models.  Indeed, \cite{GnedinJ:01} have
shown that in models in which reionisation is produced by UV from stellar
sources the OV and ``patchy reionisation'' signals are instead
strongly correlated.

The temperature anisotropies on these scales are dominated by
non-linear structures and an analytical derivation is not possible.
But it is clear from \fig~\ref{figFromNickCMB} that the early start of
IGM ionisation in the X-ray models produces additional power on arcsec
scales that should be roughly proportional to the mean fractional
ionisation of the IGM during the extended period of X-ray
pre-ionisation. In principle, the detection of anisotropies on these
scales can be used to obtain precise measurements of the ionisation
history of the IGM before the epoch of overlap when the reionisation
is competed.


\section{Discussion and Summary}\label{sec:conc}

The present paper is the third in a series (see paper~I and paper~IIa)
devoted to the study of physically plausible reionisation scenarios.
The models have optical depths to Thompson scattering \taue$\sim
0.17$, as measured by WMAP \citep{Kogut:03}, and are consistent with
observations of the IGM optical depth to \lya and \lyb photons toward
high redshift quasars \citep{Fan:03, Songaila:04}. We have studied
models of reionisation by stars (paper I) or mini-quasars (paper IIa)
considering realistic physical scenarios and observational
constraints. Our main conclusion is that it is very difficult for any
of these models to produce optical depths in excess of 0.17. If this
large optical depth is produced by ultraviolet radiation from stellar
sources (\ie, \pop3 stars with a top-heavy IMF), we find that
zero-metallicity stars must be the dominant mode of star formation up
to redshift $z\sim 10$. If this scenario is correct, we need to
understand the reasons for the very inefficient mixing of metal
enriched gas from SNe with the gas in which star formation takes
place. At the moment this inefficient mixing is not reproduced by
numerical simulations. A feasible alternative scenario requires that
most \pop3 stars must implode into black holes without exploding as SNe
or have subluminous SN explosions with a large amount of metal
fall-back onto the compact remnant in order to reduce their metal
yields. A scenario in which a large fraction of \pop3
stars end their lives as pair-instability supernovae is not compatible
with the large optical depth to Thompson scattering measured by WMAP.

In this paper we use cosmological simulations to study a reionisation
scenario in which standard reionisation by \popII stars is preceded by
partial ionisation and reheating at early times by an X-ray background
(\ie, ``X-ray preionisation'' models studied in paper~IIa using
semianalytic simulations). In these models the ionisation rate from the
secondary radiation produced by accretion on compact remnants from the
first stars dominates over the primary ultraviolet radiation emitted by
the stars during their lifetime. The most appealing aspect of these
models is their insensitivity to the duration of the epoch of
\pop3 stars domination. In fact, the relative importance of \pop3
stars with respect to \popII stars in practice has to be treated as a
free parameter due to the large uncertainties in modelling the complex
physical processes that regulate metal production and mixing.

In order to reproduce WMAP results we find that a fraction of about
$10^{-4}$ of all the baryons in the Universe needs to be accreted by
compact objects before redshift 10.  This fraction is comparable in
mass to the observed mass density of supermassive BHs in the galactic
nuclei today.  This does not pose a strong constraint on the models
because a sizeable fraction of these early intermediate-mass BHs is
expected to be removed from the galaxies during the last phases of the
merger or to reside in the interstellar medium of galaxies, being
hardly detectable \cite{MadauR:01}. Perhaps a small fraction of these
intermediate-mass BHs may be accreting and contribute to the observed
population of ULX \citep{Agol:02,Krolik:04}.

The stronger constraint on the model is posed by the observed soft
X-ray background. Assuming that the early mini-quasars population
produces an optical depth to Thompson scattering \taue$\sim 0.17$,
their contribution to the background in the 5-50 keV bands is 5-20\%.
Future X-ray missions may be able to detect these sources if they
exist. The predicted fluxes of the most rare objects is at the limit
of the detection of the Chandra deep fields and about 1000 objects
with flux $10^{-18}$ erg cm\mm s\m should be present per 5.3
arcmin$^2$.  Note that these calculations assume a spectral energy
distribution of the sources that has a non-negligible high energy
power-law component (\cf, \fig~4 in paper~IIa), in agreement with
observations of QSOs and ULX in the local universe.  If the source
spectra are dominated by a multicolour disk thermal component their
contribution to both the X-ray background in the 2-50 keV bands and
the faint source counts would be negligible.

The redshifted X-ray background also has interesting consequences for
the reionisation history of He and the thermal history of the IGM at
redshift $z \sim 3$, independently of the assumed spectra of the
sources. In this paper we confirm the results of paper~IIa in which we
found that \GII is almost fully reionised for the first time at
redshift $z \sim 17$ and afterwards slowly recombines before
experiencing a second reionisation at redshift $z \sim 3$ produced by
the redshifted X-ray background. The heating rate from the background
radiation keeps the temperature of the IGM at about 10,000 K, in rough
agreement with observations of the line widths of the \lya forest at
$z \sim 3-4$. We also emphasise that the redshifted X-ray background
is important in producing a fully ionised atomic hydrogen in the low
density intergalactic medium before stellar reionisation at redshift
$z \sim 6-7$. As a result stellar reionisation is characterised by an
almost instantaneous ``overlap phase'' of \HII regions.

The patchy topology of reionisation produced by stellar sources
contrast with the spatially homogeneous partial ionisation by X-rays.
This produces distinctive signatures on temperature/polarisation
CMB anisotropies and on the redshifted 21cm emission/absorption from
neutral hydrogen at high redshift.
\begin{enumerate}
\item 
The power spectrum of the EE polarisation is sensitive to the
visibility function $g(z)$, defined in \eq~(\ref{eq:vis}). The Plank
satellite should be able to distinguish between the visibility
function produced by an early X-ray partial ionisation (\cf,
\fig~\ref{fig:srei5} and \fig~8 in paper~IIa) or the one expected for
reionisation by stellar sources (\cf, \fig~4 in paper~I).
\item On small angular scales ($< 1 arcmin$ or $l>10^4$) the secondary
  anisotropies produced by non-linear structures during the early
  reionisation epochs dominate over other contributions, offering a
  unique opportunity to study the first episode of structure
  formation.  A remarkable feature can be observed: while in a model
  without early episode of X-ray emission the power spectrum falls off
  at small angular scales approximately as $l^{-3/4}$, in X-ray
  preionisation models the power instead grows roughly as $l^{3/4}$.
  The temperature anisotropies are of the order of $\Delta T \sim 16
  \mu$K at $\sim 1$ arcsec scales. But it is extremely difficult to
  detect anisotropies on these scales, partly because of foreground
  sources contamination \citep{Fomalont:93,Church:97}. In models in
  which reionisation is produced by UV from stellar sources the
  secondary anisotropies produced by the nonlinear Ostriker-Vishniac
  effect (\ie, assuming uniform ionisation fraction) have less power
  than in the full calculation that includes the ``patchy
  reionisation'' signal (the signals are correlated). In X-ray
  preionisation models the nonlinear Ostriker-Vishniac effect and
  ``patchy reionisation'' signals are instead slightly
  anti-correlated.
\item
The redshifted 21cm emission/absorption from neutral hydrogen is
another powerful probe of the ionisation state of the IGM at high
redshift.  In X-ray preionisation models a sharp absorption feature at
$z\sim 25-30$ falls just outside the broadcast TV bands. This feature
will appear as a roughly 100 mK ``absorption line'' in the spectrum of
the Galactic foreground, and may be observable with LOFAR. The partial
ionisation of the IGM in the X-ray preionisation scenarios offers
better opportunities to observe the fluctuations of the redshifted
21cm line in emission at redshifts lower than 10, where the signal is
easier to measure.
\end{enumerate}

Finally indirect signatures of X-ray preionisation are related to the
discovery of massive BHs in the nuclei of dwarf galaxies and/or the
identification of intermediate mass BHs in the ISM of galaxies (\eg,
ULX).  If \pop3 stars are instead responsible for the large optical
depth measured by WMAP future observations with JWST and ground based
near-infrared \lya surveys using gravitational lenses may soon be able
to observe these objects \citep{Pello:04, RicottiH:04} and probe the
ionisation state of the IGM.

\subsection*{ACKNOWLEDGEMENTS}
MR is supported by a PPARC theory grant. NG was partially supported by
by NSF grant AST-0134373 and by National Computational Science
Alliance under grant MCA03S023 and utilised IBM P690 array at the
National Center for Supercomputing Applications. Research conducted in
cooperation with Silicon Graphics/Cray Research utilising the Origin
3800 supercomputer (COSMOS) at DAMTP, Cambridge.  COSMOS is a UK-CCC
facility which is supported by HEFCE and PPARC. MR thanks Martin
Haehnelt and the European Community Research and Training Network
``The Physics of the Intergalactic Medium'' for support.  The authors
would like to thank Andrea Ferrara, Martin Haehnelt, Piero Madau and
Martin Rees for stimulating discussions.

\bibliographystyle{/home/ricotti/Latex/TeX/apj}
\bibliography{/home/ricotti/Latex/TeX/archive}

\label{lastpage}
\end{document}